# New Governing Equations for Fluid Dynamics

Chaoqun Liu[1, *], Zhining Liu[2],

[1]Department of Mathematics, University of Texas at Arlington, Texas 76019, USA

[2]HyperComp, Inc., Thousand Oaks, CA 91361, USA

## Abstract

The difference in governing equation between inviscid and viscous flows is the introduction of viscous terms. Traditional Navier-Stokes (NS) equations define the stress based on Stokes's assumptions. In NS, the stress is supposed proportional to strain and both strain and stress tensors are symmetric. There are several questions with NS, which include: 1. Both symmetric shear terms and stretching terms in strain and stress are coordinate-dependent and thus not Galilean invariant; 2. The physical meaning of both diagonal and off-diagonal elements are not clear, which is coordinate-dependent; 3. It is hard to measure the strain and stress quantitatively, and viscosity is really measured by vorticity not by symmetric strain; 4. There is no vorticity tensor in NS, which plays important role in fluid flow especially for turbulent flow. The new proposed governing equations for fluid dynamics use vorticity tensor only, which is anti-symmetric. The advantages include: 1. Both shear and stress are anti-symmetric, which are Galilean invariant and independent of coordinate rotation; 2. The physical meaning of off diagonal elements is clear, which is anti-symmetric shear stress, 3. Viscosity coefficients are obtained by experiment which uses vorticity, 4. The vorticity term can be further decomposed to rigid rotation and anti-symmetric shear, which are important to turbulence research, 5. The computation cost for viscous term is reduced to half as the diagonal terms are all zero as six elements are reduced to three. Several computational results are made, which clearly demonstrate both NS and new governing equations have exactly same results. As shown below, the new governing equation is identical to NS in mathematics, but the former has lower cost and several advantages mentioned above including possibility to better study turbulent flow. There are reasons to use the new governing equations to replace Navier-Stokes equations. The unique definition and operation of vector and tensor by matrix and matrix operation are also discussed in this paper.

## 1. Introduction

Navier-Stokes (NS) equation is a non-linear partial differential equation governing the momentum conservation for fluid flow, which is a major equation of a non-linear governing system in fluid dynamics.

Corresponding Author: Chaoqun Liu cliu@uta.edu

Daniel Bernoulli (1738) [1] and Leonhard Euler (1755) [2] subsequently derived the equation of inviscid flow which is called Euler's inviscid equations. Claude-Louis Navier (1827) [3] and others had carried out studies to explore the mathematical model of fluid flow, who considered the viscous force. In 1845, George Stokes [4] had derived the equation of motion of a viscous flow by adding Newtonian viscous terms, and finalized the Navier-Stokes Equations which have been used for almost two centuries. They were developed over several decades of progressively building the theories, from 1822 (Navier) to 1842–1850 (Stokes). NS equations are used as a common tool to solve almost all fluid flow problems everywhere. They can be used to model the weather, ocean currents, water flow in a pipe and air flow around a wing and many others. However, there are few studies to find how to understand the physical meaning of the viscous terms in NS. As well known, Stokes has three assumptions: 1. The force on fluids is the stationary pressure when the flow is stationary; 2. Fluid viscosity is isotropic; 3. Fluid flow follows Newton's law that fluid stress and strain have linear relations. These assumptions finalize the NS equation which is the major equation of our governing system in fluid dynamics [5]. Although assumption 3 does not work for non-Newtonian fluids, people in general accept these assumptions work for air, water and most common fluid flows. On the other hand, viscosity co-exists with vorticity, but the NS equation has no vorticity in its version. This is caused by Helmholtz (1857) [7] or Cauchy-Stokes [8] decomposition that the velocity gradient tensor can be decomposed to a symmetric strain part and anti-symmetric vorticity part. Because the strain part is symmetric, the stress part will be symmetric according to the Stokes's assumptions. According to recent studies (Kolar 2007[9], Li et al. [10], Liu et al. 2018, 2019 [11-12], Gao et al. 2019 [13]), the vorticity part has shear as well. The shear stress is not contributed by strain only and vorticity part must be considered as one of the sources of the stress. In this paper, vorticity part is considered as the only source of fluid stress for the purpose of computation cost reduction. In fact, the fluid shear stress is contributed by both strain and vorticity. In mathematics, the computation of stress can be performed by strain only, vorticity only, or both. The computational results are exact same. The NS equation adopts strain which is symmetric and based on Stokes's assumption. In this paper, the new governing equation adopts a new assumption which accepts that fluid stress has linear relation with vorticity which is anti-symmetric. Note that we will further modify the linear relation by considering the second order terms in future research on turbulence, but linear relation is given in the current paper for now. According to the mathematical analysis, the new governing equation is identical to NS in mathematics, but in physical sense, the new governing equation is just the opposite to NS as it assumes fluid stress is proportional to vorticity, both are anti-symmetric, which is not strain to against the Stokes's assumption and against the current NS equation.

Although both NS and the new governing equation will lead to same computational results for laminar flow, the new governing equation has several advantages: 1. The vorticity tensor is anti-symmetric which has three elements, but NS using strain tensor which has six elements. The computational cost will be reduced to half for the viscous term; 2. The anti-symmetric matrix is independent of coordinate system change or Galilean invariant, but symmetric matrix which NS uses is not; 3. The physical meaning is clear that viscous term is generated by vorticity, but not only strain; 4. The viscosity is obtained by experiment which is based on vorticity but not strain since both strain and stress are hard to measure by experiment; 5. Since vorticity can be further decomposed to rigid rotation and pure anti-symmetric shear, which will be very useful to further study turbulent flow. However, the NS equation has no vorticity term, which gives a hurdle for further turbulence research. Liu et al. (2014) [14] studied the mechanism of turbulence generation and conclude the shear instability and transformation from shear to rotation is the path of flow transition from laminar flow to turbulent flow. In Ref [15], Zhou et al. elaborated the hydrodynamic instabilities induced turbulent mixing in wide areas including inertial confinement fusion, supernovae and their transition criteria. Since the new governing equation has vorticity term which can be further decomposed to shear and rigid rotation, the new governing equation would be helpful to study on flow instability and transition.

The rest of the paper is organized as follows. Section 2 provides some mathematical background for vector and tensor operations; Section 3 describes the relation between stress and vorticity; The new governing equation is given in Section 4; Section 5 carries out several computation examples which demonstrate that the new governing equation has same computation results as NS but the computation cost is reduced; Some conclusions are made in the final section.

## 2. Mathematical background

Vector and tensor are physical quantities which are unique and coordinate-independent. However, people use matrix to represent first and second order tensors. The matrix form of vector and tensor are not unique, which depends on the definition, column or row, and coordinates.  This section tries to give a unique matrix definition for vector and tensor in fluid dynamics. Of course, the matrix is still coordinate-dependent. Meanwhile, the vector and tensor operations are different from matrix operations. This section also tries to give a unique matrix operation to represent tensor operation like tensor outer and dot operations by matrix multiplication.

### 2.1 Vector definition

In this paper, vector is defined as column matrix (not row matrix) for uniqueness. Apparently, 3D vector is unique but the column matrix is $3 \times 1$, which is not unique and related to the coordinate system.

**2.2 Tensor and matrix** [8]

Like velocity, pressure, vorticity, velocity gradient tensor is a physical quantity. The velocity gradient tensor is defined as:

$$d\vec{v} = \begin{bmatrix} \frac{\partial u}{\partial x}dx + \frac{\partial u}{\partial y}dy + \frac{\partial u}{\partial z}dz \\ \frac{\partial v}{\partial x}dx + \frac{\partial v}{\partial y}dy + \frac{\partial v}{\partial z}dz \\ \frac{\partial w}{\partial x}dx + \frac{\partial w}{\partial y}dy + \frac{\partial w}{\partial z}dz \end{bmatrix} = \begin{bmatrix} \frac{\partial u}{\partial x} & \frac{\partial u}{\partial y} & \frac{\partial u}{\partial z} \\ \frac{\partial v}{\partial x} & \frac{\partial v}{\partial y} & \frac{\partial v}{\partial z} \\ \frac{\partial w}{\partial x} & \frac{\partial w}{\partial y} & \frac{\partial w}{\partial z} \end{bmatrix} \begin{bmatrix} dx \\ dy \\ dz \end{bmatrix} = \begin{bmatrix} \frac{\partial u}{\partial x} & \frac{\partial v}{\partial x} & \frac{\partial w}{\partial x} \\ \frac{\partial u}{\partial y} & \frac{\partial v}{\partial y} & \frac{\partial w}{\partial y} \\ \frac{\partial u}{\partial z} & \frac{\partial v}{\partial z} & \frac{\partial w}{\partial z} \end{bmatrix} \cdot \begin{bmatrix} dx \\ dy \\ dz \end{bmatrix} = \nabla \vec{v} \cdot d\vec{l} \quad (1)$$

Note that velocity gradient tensor at each point is unique, but the corresponding matrix is non-unique and strongly dependent on the matrix definition and the coordinates. In addition, we must do transpose of the left matrix if we add (or remove) the dot, or $\boldsymbol{a} \cdot \boldsymbol{b} = \boldsymbol{a}^T \boldsymbol{b}$ where $\boldsymbol{a}$ is a matrix.

**2.3 Tensor operations and matrix operation** [16-18]

Outer and dot products are tensor operations. They must be transferred to algebraic operation or matrix operation. In order to get unique answer, we define vector as a column and give following definitions:

Definition 1. $\vec{a}\vec{b} = \vec{a} \otimes \vec{b} = \vec{a}(\vec{b})^T = \begin{bmatrix} a_1 \\ a_2 \\ a_3 \end{bmatrix} [b_1 \ b_2 \ b_3] = \begin{bmatrix} a_1b_1 & a_1b_2 & a_1b_3 \\ a_2b_1 & a_2b_2 & a_2b_3 \\ a_3b_1 & a_3b_2 & a_3b_3 \end{bmatrix} = \sum_{i=1}^{3}\sum_{j=1}^{3} a_i b_j \vec{j}\vec{i}$ (2)

Definition 2. $\vec{a} \cdot \vec{b} = \vec{a}^T \vec{b} = [a_1 \ a_2 \ a_3] \begin{bmatrix} b_1 \\ b_2 \\ b_3 \end{bmatrix} = \sum_1^3 a_i b_i$ (3)

Definition 3. Above definitions are valid for tensors

Tensor operations could be dot or outer, but matrix operation only has plus, minus or multiplication. We can remove $\otimes$ with transpose of the right term and remove **dot** with transpose of the left term.

**2.4 Hamilton operator**

Hamilton operator can be considered as a vector in matrix operation

$$\nabla = \begin{bmatrix} \frac{\partial}{\partial x} \\ \frac{\partial}{\partial y} \\ \frac{\partial}{\partial z} \end{bmatrix} \quad (4)$$

$$\nabla \vec{v} = \nabla \otimes \vec{v} = \begin{bmatrix} \frac{\partial}{\partial x} \\ \frac{\partial}{\partial y} \\ \frac{\partial}{\partial z} \end{bmatrix} [u \ v \ w] = \begin{bmatrix} \frac{\partial u}{\partial x} & \frac{\partial v}{\partial x} & \frac{\partial w}{\partial x} \\ \frac{\partial u}{\partial y} & \frac{\partial v}{\partial y} & \frac{\partial w}{\partial y} \\ \frac{\partial u}{\partial z} & \frac{\partial v}{\partial z} & \frac{\partial w}{\partial z} \end{bmatrix} \quad (5)$$

### 2.5 Definition of velocity gradient tensor by Hamilton operator

Gradient of a velocity vector can be described by using Hamilton operator

$$\text{Grad}\,(u\vec{\iota} + v\vec{j} + w\vec{k}) = \begin{bmatrix} \frac{\partial u}{\partial x} \\ \frac{\partial u}{\partial y} \\ \frac{\partial u}{\partial z} \end{bmatrix} \vec{\iota} + \begin{bmatrix} \frac{\partial v}{\partial x} \\ \frac{\partial v}{\partial y} \\ \frac{\partial v}{\partial z} \end{bmatrix} \vec{j} + \begin{bmatrix} \frac{\partial w}{\partial x} \\ \frac{\partial w}{\partial y} \\ \frac{\partial w}{\partial z} \end{bmatrix} \vec{k} = \begin{bmatrix} \frac{\partial u}{\partial x} & \frac{\partial v}{\partial x} & \frac{\partial w}{\partial x} \\ \frac{\partial u}{\partial y} & \frac{\partial v}{\partial y} & \frac{\partial w}{\partial y} \\ \frac{\partial u}{\partial z} & \frac{\partial v}{\partial z} & \frac{\partial w}{\partial z} \end{bmatrix} = \sum_{1}^{3} a_{ij} \vec{e_j} \vec{e_\iota} = \nabla \vec{v} \quad (6)$$

which contradicts with the definition in Wiki [19-20].

### 2.6 Relation between tensor and matrix

Tensor is a physical quantity which is unique and Galilean invariant. A second order tensor can be expressed by a $3 \times 3$ matrix which is not unique, but relied on the definition and coordinate systems. Therefore, we must use a unique definition and unique coordinate to represent the tensor

$$\text{Grad}\,\vec{v} = \nabla\vec{v} = \begin{bmatrix} \frac{\partial u}{\partial x} & \frac{\partial v}{\partial x} & \frac{\partial w}{\partial x} \\ \frac{\partial u}{\partial y} & \frac{\partial v}{\partial y} & \frac{\partial w}{\partial y} \\ \frac{\partial u}{\partial z} & \frac{\partial v}{\partial z} & \frac{\partial w}{\partial z} \end{bmatrix} \quad (7)$$

$$(\text{Grad}\,\vec{v})^T = (\nabla\vec{v})^T = \begin{bmatrix} \frac{\partial u}{\partial x} & \frac{\partial u}{\partial y} & \frac{\partial u}{\partial z} \\ \frac{\partial v}{\partial x} & \frac{\partial v}{\partial y} & \frac{\partial v}{\partial z} \\ \frac{\partial w}{\partial x} & \frac{\partial w}{\partial y} & \frac{\partial w}{\partial z} \end{bmatrix} \quad (8)$$

Note that the rows and columns are changed in the above matrices.

### 2.7 Cauchy-Stokes decomposition

$$(\nabla\vec{v})^T = \frac{1}{2}[(\nabla\vec{v})^T + \nabla\vec{v}] + \frac{1}{2}[(\nabla\vec{v})^T - \nabla\vec{v}] \quad (9)$$

$$\nabla\vec{v} = \frac{1}{2}[\nabla\vec{v} + (\nabla\vec{v})^T] + \frac{1}{2}[\nabla\vec{v} - (\nabla\vec{v})^T] \quad (10)$$

### 2.8 Divergence of velocity gradient tensor and its transpose

Let us define

$$tr = \frac{\partial u}{\partial x} + \frac{\partial v}{\partial y} + \frac{\partial w}{\partial z} = Tr(\nabla \vec{v}) = Tr[(\nabla \vec{v})^T] = \nabla \cdot \vec{v} \tag{11}$$

For incompressible flow, *tr=0*. Let us discuss the incompressible flow first, and compressible flow is similar.

$$(\nabla^2 \vec{v})^T = (\nabla \cdot \nabla \otimes \vec{v})^T = (\nabla \cdot \nabla [u \ v \ w])^T = \begin{bmatrix} \nabla^2 u \\ \nabla^2 v \\ \nabla^2 w \end{bmatrix} \tag{12}$$

If use tensor operation for Hamilton operator, we should have

$$(\nabla^2 \vec{v})^T = (\nabla \cdot \nabla \otimes \vec{v})^T = ([\frac{\partial}{\partial x}, \frac{\partial}{\partial y}, \frac{\partial}{\partial z}] \begin{bmatrix} \frac{\partial u}{\partial x} & \frac{\partial v}{\partial x} & \frac{\partial w}{\partial x} \\ \frac{\partial u}{\partial y} & \frac{\partial v}{\partial y} & \frac{\partial w}{\partial y} \\ \frac{\partial u}{\partial z} & \frac{\partial v}{\partial z} & \frac{\partial w}{\partial z} \end{bmatrix})^T = [\frac{\partial^2 u}{\partial x^2} + \frac{\partial^2 u}{\partial y^2} + \frac{\partial^2 u}{\partial z^2}, \frac{\partial^2 v}{\partial x^2} + \frac{\partial^2 v}{\partial y^2} + \frac{\partial^2 v}{\partial z^2},$$

$$\frac{\partial^2 w}{\partial x^2} + \frac{\partial^2 w}{\partial y^2} + \frac{\partial^2 w}{\partial z^2}]^T = \begin{bmatrix} \frac{\partial^2 u}{\partial x^2} + \frac{\partial^2 u}{\partial y^2} + \frac{\partial^2 u}{\partial z^2} \\ \frac{\partial^2 v}{\partial x^2} + \frac{\partial^2 v}{\partial y^2} + \frac{\partial^2 v}{\partial z^2} \\ \frac{\partial^2 w}{\partial x^2} + \frac{\partial^2 w}{\partial y^2} + \frac{\partial^2 w}{\partial z^2} \end{bmatrix} = \begin{bmatrix} \nabla^2 u \\ \nabla^2 v \\ \nabla^2 w \end{bmatrix} \neq \nabla \cdot \nabla \otimes \vec{v} = \nabla^2 \vec{v} \tag{13}$$

Each time when we use the Hamilton operator, we must do the transpose.

For incompressible flow,
$$\frac{\partial u}{\partial x} + \frac{\partial v}{\partial y} + \frac{\partial w}{\partial z} = 0$$

$$[\nabla \cdot (\nabla \vec{v})^T]^T = ([\frac{\partial}{\partial x}, \frac{\partial}{\partial y}, \frac{\partial}{\partial z}] \begin{bmatrix} \frac{\partial u}{\partial x} & \frac{\partial u}{\partial y} & \frac{\partial u}{\partial z} \\ \frac{\partial v}{\partial x} & \frac{\partial v}{\partial y} & \frac{\partial v}{\partial z} \\ \frac{\partial w}{\partial x} & \frac{\partial w}{\partial y} & \frac{\partial w}{\partial z} \end{bmatrix})^T = \begin{bmatrix} \frac{\partial^2 u}{\partial x^2} + \frac{\partial^2 v}{\partial y \partial x} + \frac{\partial^2 w}{\partial z \partial x} \\ \frac{\partial^2 u}{\partial x \partial y} + \frac{\partial^2 v}{\partial y^2} + \frac{\partial^2 w}{\partial z \partial y} \\ \frac{\partial^2 u}{\partial x \partial z} + \frac{\partial^2 v}{\partial y \partial z} + \frac{\partial^2 w}{\partial z^2} \end{bmatrix} = \begin{bmatrix} \frac{\partial}{\partial x}(\frac{\partial u}{\partial x} + \frac{\partial v}{\partial y} + \frac{\partial w}{\partial z}) \\ \frac{\partial}{\partial y}(\frac{\partial u}{\partial x} + \frac{\partial v}{\partial y} + \frac{\partial w}{\partial z}) \\ \frac{\partial}{\partial z}(\frac{\partial u}{\partial x} + \frac{\partial v}{\partial y} + \frac{\partial w}{\partial z}) \end{bmatrix} = 0$$

(14)

For compressible flow

$$[\nabla \cdot (\nabla \vec{v})^T]^T = \begin{bmatrix} \frac{\partial}{\partial x} tr \\ \frac{\partial}{\partial y} tr \\ \frac{\partial}{\partial z} tr \end{bmatrix} = \nabla tr = \nabla(\nabla \cdot \vec{v}) \tag{15}$$

The conclusion is that the divergence of velocity gradient tensor transpose $[\nabla \cdot (\nabla \vec{v})^T]^T$ is $\nabla(\nabla \cdot \vec{v})$ for compressible flow and is ZERO for incompressible flow. However, the divergence of the velocity gradient tensor or $[\nabla \cdot \nabla \vec{v}]^T$ is not zero.

It is suggested to us $\begin{bmatrix} \nabla^2 u \\ \nabla^2 v \\ \nabla^2 w \end{bmatrix} = [\nabla \cdot \nabla \vec{v}]^T$. In the N-S equations, we have demonstrated that

$[\nabla \cdot (\nabla \vec{v})^T]^T = 0$ for incompressible flow.

3. **New governing equation**

   **3.1 Navier-Stokes equations**

   Newton's second law gives

   $$\vec{F} = ma$$

   where $\vec{F}$ is the force on a particle, $m$ is the mass of the particle and $a$ is the acceleration of the particle. If we use the vector form:

   $$\rho \frac{d\vec{v}}{dt} = \nabla \cdot \mathbf{F} + \rho \vec{f}$$

   or

   $$\frac{\partial(\rho \vec{v})}{\partial t} + \nabla \cdot (\rho \vec{v} \vec{v}) = \rho \vec{f} - \nabla p - \frac{2}{3} \nabla[\mu(\nabla \cdot \vec{v})] + \{\nabla \cdot [\mu(\nabla \vec{v} + (\nabla \vec{v})^T)]\}^T \quad (16)$$

   since

   $$\mathbf{F} = \{-p - \frac{2}{3}[\mu(\nabla \cdot \vec{v})]\}\mathbf{I} + 2\mu \mathbf{A}, \quad \mathbf{A} = \frac{1}{2}(\nabla \vec{v} + \nabla \vec{v}^T),$$

   where $\mathbf{F}$ is the force tensor

   **3.2 Velocity gradient tensor CS decomposition**

$$\nabla \vec{v} = \frac{1}{2}[\nabla \vec{v} + (\nabla \vec{v})^T] + \frac{1}{2}[\nabla \vec{v} - (\nabla \vec{v})^T] = T_S + T_V \quad (17)$$

which are called strain tensor and vorticity tensor. Let us assume the stress tensor:

$$\mathbf{F} = \frac{1}{2}\mu[\nabla \vec{v} + (\nabla \vec{v})^T] + \frac{1}{2}\mu[\nabla \vec{v} - (\nabla \vec{v})^T] \quad (18)$$

$$(\nabla \cdot \mathbf{F})^T = (\nabla \cdot \{\frac{1}{2}\mu[\nabla \vec{v} + (\nabla \vec{v})^T] + \frac{1}{2}\mu[\nabla \vec{v} - (\nabla \vec{v})^T]\})^T \quad (19)$$

Assume $\nu$ is constant, because

$$\nabla \cdot (\nabla \vec{v})^T = 0$$

for incompressible flow. It has same form as

$$\nabla \cdot \mathbf{F} = \nabla \cdot (2\mu) \frac{1}{2}[\nabla \vec{v} + (\nabla \vec{v})^T]$$

which has been misunderstood as the stress is produced by strain only, and no contributions are from vorticity tensor.

### 3.3 New assumption for shear and stress

In physics, the stress is contributed equally by the strain tensor and vorticity tensor. In mathematics, use of strain or vorticity to represent is identical. For clear physical meaning, reduction of computation cost, and deep turbulence understanding, the new governing equation uses the vorticity to calculate the stress and assumes both shear and stress are anti-symmetric which means: 1. Shear is anti-symmetric; 2. Stress is proportional to shear; 3. Stress is anti-symmetric. These assumptions are just the opposite to Stokes's assumption.

We can also write

$$\nabla \cdot \boldsymbol{F} = \nabla \cdot (2\,\mu)\frac{1}{2}[\nabla\vec{v} - (\nabla\vec{v})^T] = \nabla \cdot \mu[\nabla\vec{v} - (\nabla\vec{v})^T], \tag{20}$$

which is anti-symmetric. Use of vorticity tensor has several advantages since the vorticity tensor has three independent elements only without diagonal elements. All three elements are components of vorticity which is Galilean invariant.

$$[\nabla\vec{v} - (\nabla\vec{v})^T] = \begin{bmatrix} 0 & \omega_z & -\omega_y \\ -\omega_z & 0 & \omega_x \\ \omega_y & -\omega_x & 0 \end{bmatrix} \tag{21}$$

We will see that it is much better to use the vorticity tensor to derive the governing equations. It also makes sense that the viscosity is generated from vorticity not from strain since no one can easily measure strain and stress. All the measurements made by experiment are using vorticity, i.e.

$$\text{F} = \left(\frac{\mu}{\rho}\right)\frac{\partial u}{\partial y} = -2\left(\frac{\mu}{\rho}\right)\omega_z \text{ or } \mu = \rho F/(-2\omega_z), \tag{22}$$

where F is the magnitude of the force driving the plate (Figure 1).

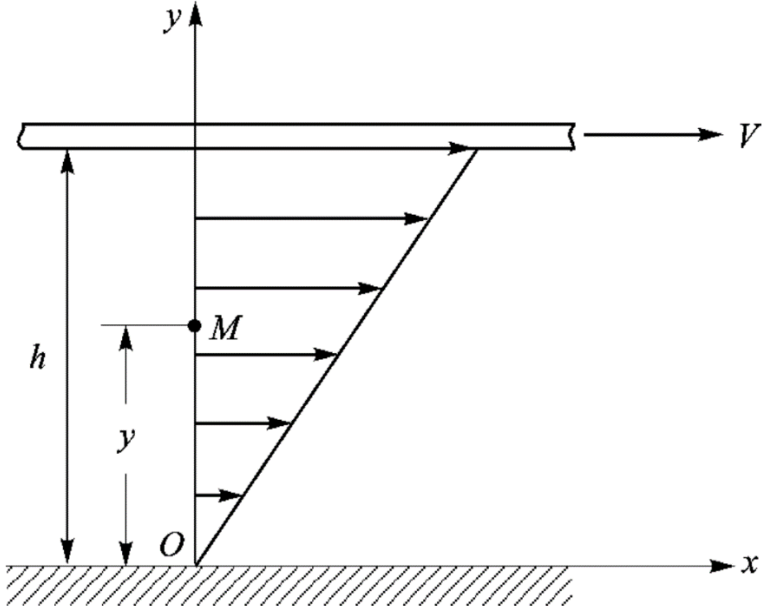

Figure 1 Shear force $F = \mu/\rho \cdot \frac{\partial u}{\partial y}$

The original Navier-Stokes equations can be written as

$$\frac{\partial(\rho\vec{v})}{\partial t} + \nabla \cdot (\rho\vec{v}\vec{v}) = \rho\vec{f} - \nabla p - \frac{2}{3}\nabla[\mu(\nabla \cdot \vec{v})] + \{\nabla \cdot [\mu(\nabla\vec{v} + (\nabla\vec{v})^T)]\}^T, \quad (23)$$

which assumes the strain and stress are symmetric and six elements have been used in the stress tensor, which are coordinate-dependent. If we use vorticity tensor which only has three independent elements with all diagonal elements equal to zero. The three components of vorticity are Galilean invariant and coordinate-independent.

As we addressed above, since $\nabla \cdot (\nabla\vec{v})^T = \nabla(\nabla \cdot \vec{v})$, we will have a new governing equation,

$$\frac{\partial(\rho\vec{v})}{\partial t} + \nabla \cdot (\rho\vec{v}\vec{v}) = \rho\vec{f} - \nabla p + \frac{4}{3}\nabla[\mu(\nabla \cdot \vec{v})] + \{\nabla \cdot [\mu(\nabla\vec{v} - (\nabla\vec{v})^T)]\}^T, \quad (24)$$

It looks like similar to NS, but the viscous terms are changed from symmetric to anti-symmetric and six elements become three elements:

$$(\nabla\vec{v} - (\nabla\vec{v})^T) = \begin{bmatrix} 0 & \omega_z & -\omega_y \\ -\omega_z & 0 & \omega_x \\ \omega_y & -\omega_x & 0 \end{bmatrix} = \begin{bmatrix} 0 & \frac{\partial v}{\partial x} - \frac{\partial u}{\partial y} & -(\frac{\partial u}{\partial z} - \frac{\partial w}{\partial x}) \\ -(\frac{\partial v}{\partial x} - \frac{\partial u}{\partial y}) & 0 & \frac{\partial w}{\partial y} - \frac{\partial v}{\partial z} \\ (\frac{\partial u}{\partial z} - \frac{\partial w}{\partial x}) & -(\frac{\partial w}{\partial y} - \frac{\partial v}{\partial z}) & 0 \end{bmatrix}, \quad (25)$$

which is the vorticity tensor with three independent elements at the off-diagonal position and all zero in the diagonal position. Here, $\omega$ is vorticity which is Galilean invariant.

The stresses in the original NS equation is

$$\tau_{ij} = \mu\left(\frac{\partial u_i}{\partial x_j} + \frac{\partial u_j}{\partial x_i}\right) - \frac{2}{3}\mu\delta_{ij}\frac{\partial u_k}{\partial x_k}, \quad i,j,k = 1,2,3, \tag{26}$$

with six independent elements because it is symmetric

In the new governing equation,

$$\tau_{ij} = \mu\left(\frac{\partial u_i}{\partial x_j} - \frac{\partial u_j}{\partial x_i}\right) + \frac{4}{3}\mu\delta_{ij}\frac{\partial u_k}{\partial x_k}, \quad i,j,k = 1,2,3, i \neq j, \tag{27}$$

with three independent elements because it s anti-symmetric

### 3.6 Incompressible flow

For incompressible flow,

$$\nabla \cdot (\nabla \vec{v})^T = \nabla(\nabla \cdot \vec{v}) = 0,$$

The original NS equation is

$$\frac{\partial(\rho\vec{v})}{\partial t} + \nabla \cdot (\rho\vec{v}\vec{v}) = \rho\vec{f} - \nabla p + \{\nabla \cdot [\mu(\nabla\vec{v} + (\nabla\vec{v})^T)]\}^T \tag{28}$$

with

$$\tau_{ij} = \mu\left(\frac{\partial u_i}{\partial x_j} + \frac{\partial u_j}{\partial x_i}\right), i,j = 1,2,3,$$

which has totally six entries to be calculated. The new governing equation is written as

$$\frac{\partial(\rho\vec{v})}{\partial t} + \nabla \cdot (\rho\vec{v}\vec{v}) = \rho\vec{f} - \nabla p + \{\nabla \cdot [\mu(\nabla\vec{v} - (\nabla\vec{v})^T)]\}^T \tag{29}$$

with

$$\tau_{ij} = \mu\left(\frac{\partial u_i}{\partial x_j} - \frac{\partial u_j}{\partial x_i}\right) \ i,j = 1-3 \ and \ i \neq j,$$

which has totally three entries to be calculated.

Note that, because in the new governing equation,

$$\tau_{ii} = \mu\left(\frac{\partial u_i}{\partial x_i} - \frac{\partial u_i}{\partial x_i}\right) = 0, \tag{30}$$

all diagonal terms are voided and the six elements in computation become three. The only thing which is different between Euler equation and the NS equation is the viscous terms which Euler equation does not have. We cannot change the time term, convection term and body force term, which are all from mathematics. Now, in the new governing equations, the viscous term cost is reduced by half. More important, the vorticity term can be further decomposed to rotation and anti-shear terms which may further be useful for turbulence research. However, it is beyond the range of discussion in the current paper.

### 3.7 New governing equations in xyz coordinate

Because the new governing equation has the vector form of

$$\frac{\partial(\rho\vec{v})}{\partial t} + \nabla \cdot (\rho\vec{v}\vec{v}) = \rho\vec{f} - \nabla p + \frac{4}{3}\nabla[\mu(\nabla \cdot \vec{v})] + \{\nabla \cdot [\mu(\nabla\vec{v} - (\nabla\vec{v})^T)]\}^T, \tag{31}$$

it is easy to write in xyz-coordinates:

$$\frac{\partial \rho u}{\partial t} + \frac{\partial \rho uu}{\partial x} + \frac{\partial \rho uv}{\partial y} + \frac{\partial \rho uw}{\partial z} = \rho f_x - \frac{\partial p}{\partial x} + \frac{4}{3}\frac{\partial}{\partial x}[\mu(\nabla \cdot \vec{v})] + \left(-\frac{\partial}{\partial y}\mu\omega_z + \frac{\partial}{\partial z}\mu\omega_y\right),$$

$$\frac{\partial \rho v}{\partial t} + \frac{\partial \rho uv}{\partial x} + \frac{\partial \rho vv}{\partial y} + \frac{\partial \rho vw}{\partial z} = \rho f_y - \frac{\partial p}{\partial y} + \frac{4}{3}\frac{\partial}{\partial y}[\mu(\nabla \cdot \vec{v})] + \left(\frac{\partial}{\partial x}\mu\omega_z - \frac{\partial}{\partial z}\mu\omega_x\right),$$

$$\frac{\partial \rho w}{\partial t} + \frac{\partial \rho uw}{\partial x} + \frac{\partial \rho vw}{\partial y} + \frac{\partial \rho ww}{\partial z} = \rho f_z - \frac{\partial p}{\partial z} + \frac{4}{3}\frac{\partial}{\partial z}[\mu(\nabla \cdot \vec{v})] + \left(-\frac{\partial}{\partial x}\mu\omega_y + \frac{\partial}{\partial y}\mu\omega_x\right), \quad (32)$$

where

$$\omega_x = \frac{\partial w}{\partial y} - \frac{\partial v}{\partial z}, \omega_y = \frac{\partial u}{\partial z} - \frac{\partial w}{\partial x}, \omega_z = \frac{\partial v}{\partial x} - \frac{\partial u}{\partial y}$$

Apparently, the viscous term is reduced to half as $(\nabla \vec{v} + (\nabla \vec{v})^T)$ need to calculate 6 elements but $(\nabla \vec{v} - (\nabla \vec{v})^T)$ need only three.

### 3.8 Answers to some concerns

3.8.1 Question # 1: Stress is only related strain and both are symmetric. Therefore, vorticity or vorticity tensor did not play any role in generation of stress.

Actually, since $\nabla \cdot (\nabla \vec{v})^T = 0$ for incompressible flow, but $\nabla \cdot \nabla \vec{v} \neq 0$,

$$2\nabla \cdot \frac{1}{2}[\nabla \vec{v} + (\nabla \vec{v})^T] = 2\nabla \cdot \frac{1}{2}[\nabla \vec{v} - (\nabla \vec{v})^T] \quad (33)$$

For compressible flow

$$\nabla \cdot [\nabla \vec{v} + (\nabla \vec{v})^T] = \nabla \cdot [\nabla \vec{v} - (\nabla \vec{v})^T] + 2\nabla(\nabla \cdot \vec{v}) \quad (34)$$

The symmetric tensor and anti-symmetric tensor play same role and thus strain and vorticity play same role. We can either use the strain or the vorticity or both to derive governing equations for fluid flow. However, use of vorticity tensor is simpler for laminar flow and likely much more useful for turbulent flow.

3.8.2 Question # 2: Strain and stress are symmetric

Let us take a 2D square as an example,

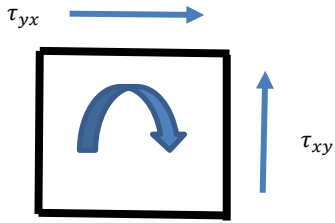

Fig. 2. 2D element

Many research papers and books use the angular momentum conservation (see Fig. 2). Note that Stokes ignored the contribution of force generated by vorticity. This is from the order analysis of the angular momentum conservation, and the fluid particle is infinitesimal. For a 2-D square (Fig. 2), the angular momentum balance can be written:

$$\sqrt{\Delta x^2 + \Delta y^2}\rho\Delta x\Delta y\dot{\theta} = \tau_{yx}\Delta x\Delta y - \tau_{xy}\Delta y\Delta x \tag{35}$$

For an infinitesimal particle
$\sqrt{\Delta x^2 + \Delta y^2}\Delta x\Delta y \ll \Delta x\Delta y$,
the left-hand side has high order, and then $\tau_{yx} = \tau_{xy}$ to conclude the stress-rate tensor is symmetric from dimensional analysis,

$$\tau_{yx}\Delta x\Delta y - \tau_{xy}\Delta y\Delta x \approx 0.$$
$$\tau_{yx} = \tau_{xy} \tag{36}$$

However, this conclusion and derivation is made by one side and the correct analysis of angular momentum conservation should consider both sides and can be described as follows.

Let us take a 2D square as an example (Figure 3),

$$\sqrt{\Delta x^2 + \Delta y^2}\rho\Delta x\Delta y\dot{\theta} = \frac{\partial}{\partial y}\tau_{yx}\Delta x\Delta x\Delta y - \frac{\partial}{\partial x}\tau_{xy}\Delta x\Delta y\Delta x \tag{37}$$

Assume
$$\Delta x = \Delta y$$
$$\sqrt{2}\rho\dot{\theta} = \frac{\partial}{\partial y}\tau_{yx} - \frac{\partial}{\partial x}\tau_{xy} \neq 0 \tag{38}$$

We can clearly see the rotation is not caused by shear stress, but by the gradient of shear stress. It is questionable to assume strain and stress are symmetric. In that way, the stress produced by vorticity is ignored.

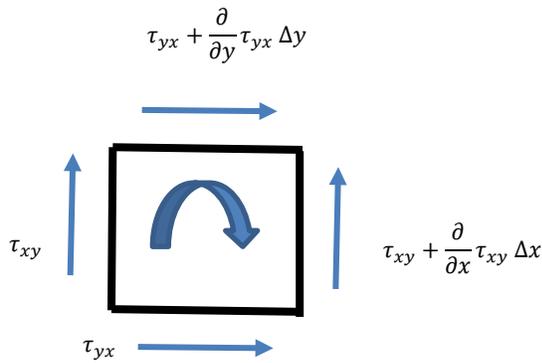

Fig 3. 2D element with shear

## 4. Computational examples

The new form of the governing equation was implemented in HyPerComp in-house compressible unstructured finite volume code HYCE for verifications. Numerically, we only need to change the viscous flux related terms in momentum and energy equations. Compressible form of the equations was also derived to extend the applications to higher Mach number flows.

### 4.1 Backward facing step with expansion ratio 1.942

The first verification case is a laminar flow case: a 2-D laminar flow over a backward-facing step at various Reynolds numbers. Here, the inflow Mach No. was set at M=0.01 so the flow compressibility influence can be minimized. Fig.4 depicts the computational domain setup, with the channel expansion ratio 1.942. The inlet channel was extended to 40h so the fully developed channel flow can be reached for a wide range of Reynolds numbers. We are focusing on the main recirculation bubble size, so the total domain length after the step is set at 95h, and a uniform back pressure boundary condition is imposed.

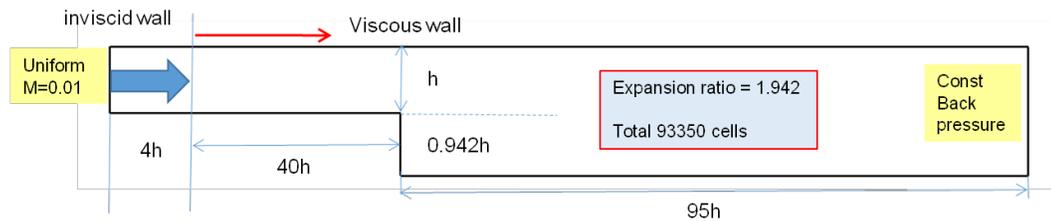

Fig.4. The backward-facing step verification case setup.

800. It is shown that the solutions obtained from both form of the governing equations (original NS and new governing equation) are basically identical, and the comparison with experimental results is quite well. In Fig.5, the main recirculation bubble length normalized by the step height $b=0.942h$ vs. Reynolds number $Re_{Dh}$ was plotted, together with the experimental data by Armaly et al.[21].

Fig. 6 gives some stream line plots for Reynolds number $100 \leq Re_{Dh} \leq 800$ obtained with the new equations. The background is pressure contours. This plot clearly shows the formation of the recirculation regions as the Reynolds number changes. Since the main purpose of this paper is to demonstrate the new equation provides the same answer as the original ones, we will not discuss the physical details about this problem. Fig. 7 compares the skin friction coefficients obtained using both equations for both bottom and top walls at $Re_{Dh}=600$. No observable difference can be found.

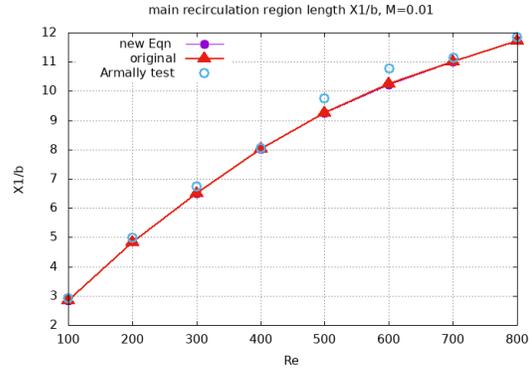

Fig. 5. Comparisons of the main recirculation region length.

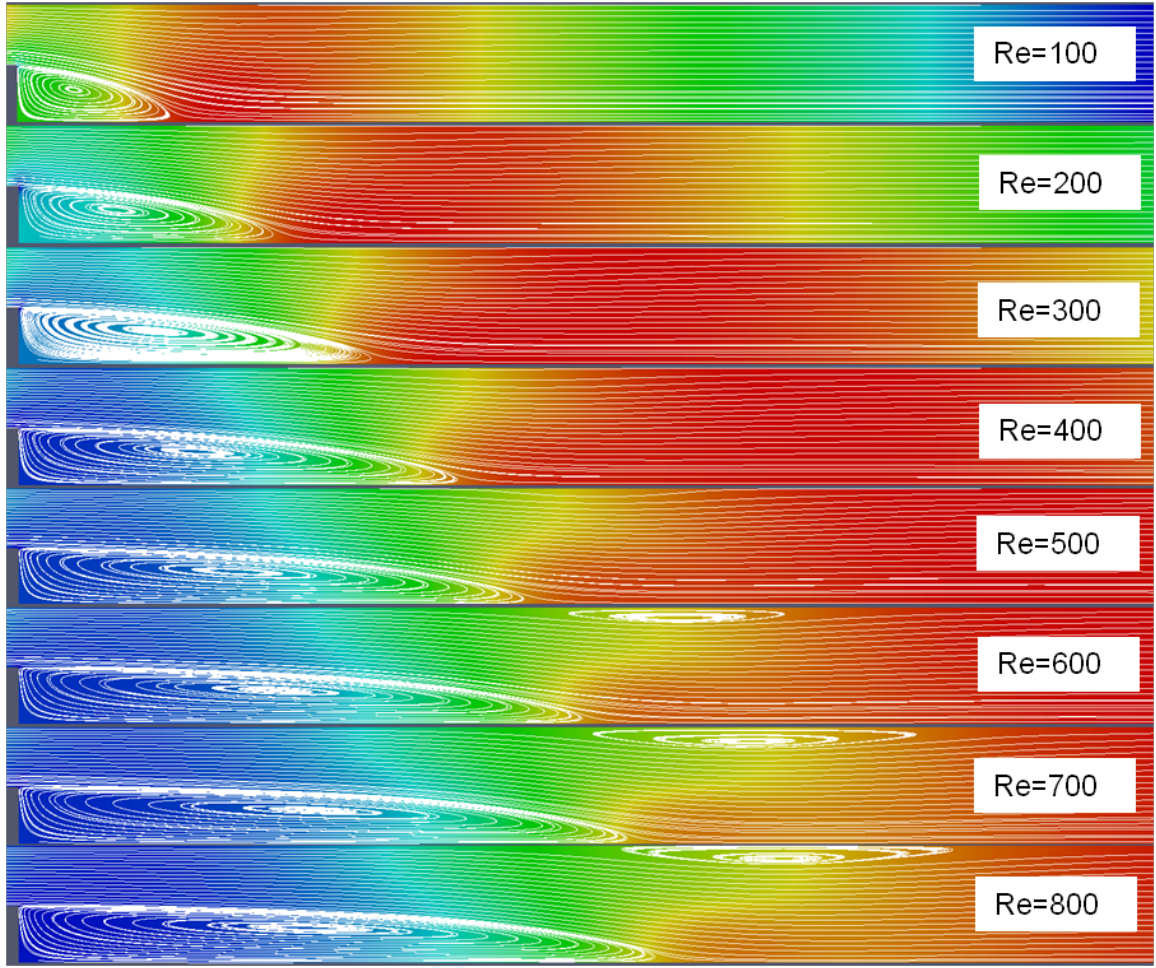

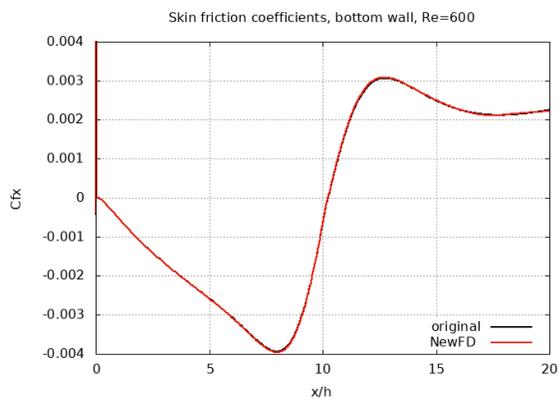 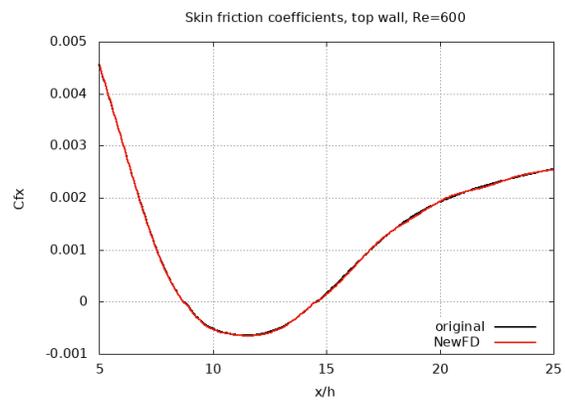

### 4.2 Flow in an S-Duct

The 2$^{nd}$ verification case is the S-shaped M2129 intake case. The flow in this intake model is quite complex. The flow accelerates into the intake from a stagnation point on the outer cowl surface, then further accelerates around the starboard side first bend of the intake. Fig. 7 gives the computational domain and problem setups. The duct geometry was truncated at engine face. A fixed mass flow condition is imposed at the exit boundary face. Inflow is set at M=0.21.

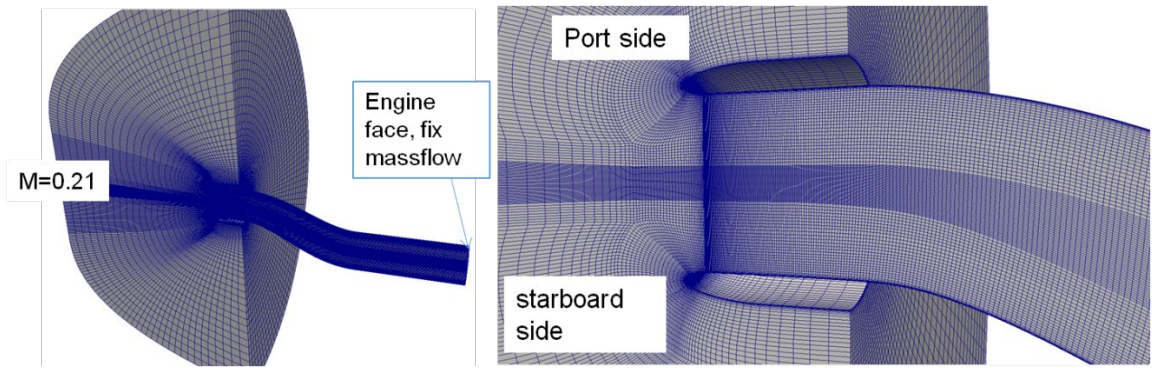

Fig.7. The M2129 intake computational domain surface mesh and intake side definitions.

Though inflow speed is low @ M=0.21, flow inside the duct can go supersonic for the high mass flow rate case. In this calculation, the allspeed scheme which is a blending of modified Rusanov and HLLE algorithms is chosen. The mesh's first point off the wall distance is ~0.005mm, resulting to y+ O(1.0). An all hex mesh with a total number of 1.54M cells for a half of the geometry was generated. At this condition, the flow is turbulent, so the turbulence models have to be used during the run. Here we choose Uri's one-equation and Menter's k-ω SST two- equation turbulence models were used. Two cases, namely "low mass rate" and "high mass rate" cases, were calculated with both the original and the new governing equations for verification. For details of the problem, refer to [22]. Here, $M_\infty$=0.21, $P_\infty$=101957.227 Pa, $\rho_\infty$=1.220434 kg/m$^3$ , $P_{total}$=105139.5 Pa. Diameter of the capture area=0.144m, capture area=0.01628686m$^2$, Mass$_\infty$=1.427535 kg/s.

For the low mass rate case, mass flow rate MFR=1.457, so the actual mass being sucked into the S-duct is 2.080 kg/s. The flow is full subsonic, both turbulence models gave quite satisfied solutions, which are also comparable to AGARD CFD solution [22]. New equation and original equations results differ slightly with turbulence model turned on, which needs further investigation in the future. Fig.8. shows the pressure trace from the starboard side of the duct for both models compared against AGARD CFD and experiment data. Flow acceleration around the cowl is well predicted. The subsequent pressure recovery is overpredicted by the Uri's 1-eqn model, while with the SST model it shows better matching. Static pressure and Mach number contours on the domain surface obtained with the new equations using SST model are given in Fig.9, showing the recirculation region in the pipe bend regime, and sharp pressure drop after inlet cowl.

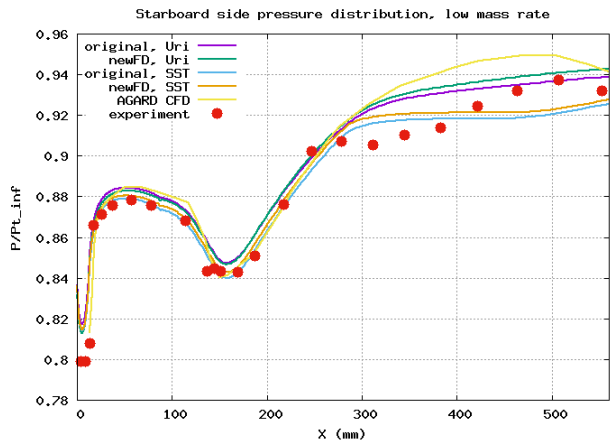

Fig. 8. Pressure extraction's from the starboard side of the intake for the low MFR case.

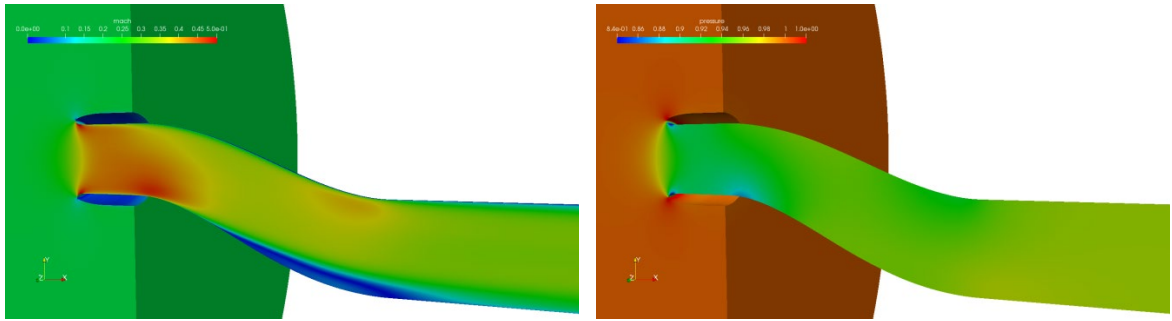

Fig.9. Mach No. (left) and static pressure (right) contours on the domain surface for the low MFR case.

For the high MFR case, MFR=2.173, mass flow is 3.1020 kg/s. This is a much more challenging problem due to the flow after the cowl becomes supersonic. The starboard side flow at the first bend is also supersonic. Incorrect prediction of the intake boundary layer can largely change the flow structure and can even choke the flow. To examine the predicted results accuracy, pressure extracted from the starboard side of the intake was again compared with the test data and AGARD CFD result, and the agreement is quite well, see Fig.10, with Uri's model predicted better pressure recovery. It can be seen that flow acceleration into the duct is over-predicted with much lower pressure compared to the data. At the mean time, it also shows the HYCE code solutions obtained with both original and new governing equations are very close, indicates the compressible form of the equation is valid. Note that supersonic flow is generated at the cowl as the flow accelerates into the duct, as well as on the starboard side at the first bend, as shown in Fig.11 (left). A complex shock reflection system in the cowl region can also be observed. Pressure contour is also shown in Fig.11(right).

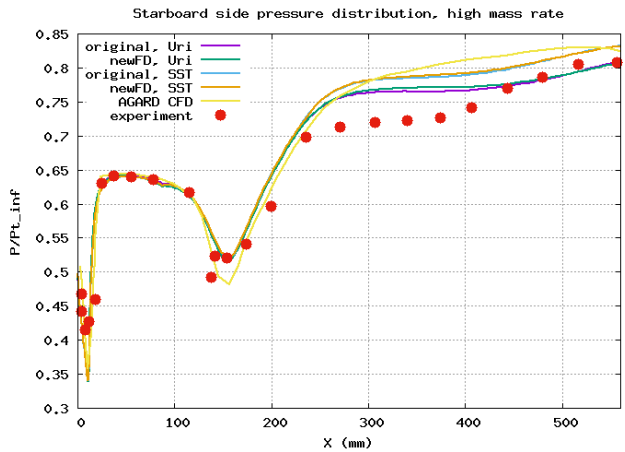

Fig. 10. Pressure extraction's from the starboard side.

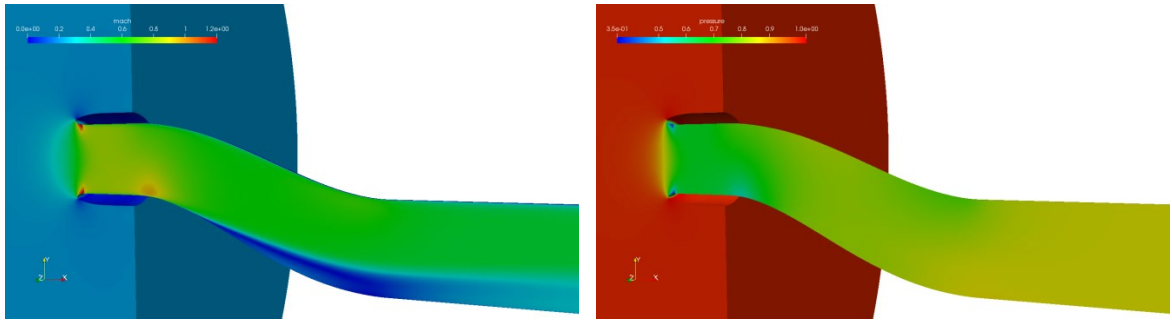

Fig.11. Mach No. (left) and static pressure (right) contours on the domain surface for the high MFR case.

### 5. Conclusions and future research

Based on the above discussion and computation, following conclusions can be made

1) Unique definition of vector and tensor is tried to be given by matrix. Unique vector and tensor operation is tried to be given by matrix multiplication.

2) The new governing equations can get same computational results for fluid flow as the Navier-Stokes equation gives because the NS and the new governing equation is identical in mathematics.

3) The new governing equations only use the vorticity tensor which is anti-symmetric. Because the anti-symmetric tensor has zeros in all diagonal elements, the computation cost for the viscous term will be reduced to half in comparison with NS equations

4) The anti-symmetric tensor which represents vorticity which is Galilean invariant. This will make coordinate transformation very easy

5) It is very hard to measure strain and stress in fluid flow. The viscosity is really obtained by experiment based on vorticity. The new governing equations coincide with the experiment with clear physical meaning that viscosity is generated by anti-symmetric shear.

6) In Navier-Stokes equations, there are symmetric shear and stretching (compression), which are both coordinate-related, not Galilean invariant, and hard to measure.

Since turbulence has many vortices which are related to vorticity tensor. Unlike Navier-Stokes equation which has no anti-symmetric vorticity tensor, the new governing equation has vorticity tensor which could be further decomposed to rotational part and anti-symmetric shear. This may be useful to research on turbulence. This will be included in the future work.

**Data availability**

The data that supports the findings of this study are available from the corresponding author upon reasonable request.

**Acknowledgement**

The first author is grateful to University of Texas at Arlington (UTA) for long time support as a tenured faculty. The first author also thanks Texas Advanced Computing Center (TACC) for long term support in provision of computation hours.